\documentclass
[twocolumn,showpacs,preprintnumbers,superscriptaddress,nofootinbib,byrevtex,nobalancelastpage]{revtex4}%
\usepackage[dvips]{graphicx}
\usepackage{amsmath}
\usepackage{bm}
\usepackage{amsfonts}
\usepackage{amssymb}
\usepackage[normal]{caption2}%
\begin{document}
\preprint{APS/123-QED}
\title{QED radiative corrections for elastic $e(\mu)p$ scattering in hadronic variables}
\author{A. A. Akhundov }
\email{akhundov@ksu.edu.sa}
\affiliation{Department of Physics and Astronomy, College of Science, King Saud University,
P.O. Box 2455, Riyadh 11454, Saudi Arabia}
\author{H. H. Alharbi}
\email{alharbi@kacst.edu.sa}
\affiliation{National Center for Mathematics and Physics, KACST, P.O. Box 6086, Riyadh
11442, Saudi Arabia }
\author{H. A. Alhendi$^{1}$}
\email{alhendi@ksu.edu.sa}
\date{\today }

\begin{abstract}
{\small A numerical analysis of QED radiative corrections for elastic
$e(\mu)p$ scattering in hadronic variables at energies of the current
experiment at JLab is performed. The explicit formulae from the review of
Akhundov et al. resulting from the integration over the phase space of
leptonic variables plus photon are used to obtain the values of the cross
sections and the radiative correction factor for unpolarized lepton-proton
scattering. Our numerical results agree with the corresponding results arising
from the formulae of Afanasev et al. }

\end{abstract}

\pacs{Valid PACS appear here}
\maketitle





{\normalsize The new detector generation allows to measure both scattered
electron and hadronic final state in deep inelastic $ep$ scattering at
HERA~\cite{HERA}
\begin{equation}
e+p\rightarrow e+X,\label{deepborn}%
\end{equation}
and elastic $ep$ scattering at the Thomas Jefferson National Accelerator
Facility (JLab)~\cite{JLab}
\begin{equation}
e+p\rightarrow e+p.\label{elborn}%
\end{equation}
}

{\normalsize The possibilities of these detectors have opened a new page in
the field of the Radiative Corrections for electron-proton
collisions\cite{Tsai,Maxi}. Physical analysis of the $ep$ collisions is based
now not only on the familiar leptonic variables, when only scattered electrons
are detected, but also on kinematical variables from the hadron measurement,
or some combinations of both, such as mixed variables\cite{Bent,Blum}. This
has to be met with a new treatment of the radiative corrections for elastic
and deep inelastic scattering in different variables. Different choices of
variables have no influence on the cross sections (\ref{deepborn}) and
(\ref{elborn}) in the Born approximation, but make huge differences in the
predictions for the radiative corrections to these processes, because the
kinematics of the bremsstrahlung contributions to the processes
(\ref{deepborn}) and (\ref{elborn}) with non-observed photon(s):}%
\newline
\begin{subequations}
{\normalsize
\begin{equation}
e+p\rightarrow e+X+n\gamma,\label{deebBrm}%
\end{equation}
}\newline and{\normalsize
\begin{equation}
e+p\rightarrow e+p+n\gamma,\tag{3b}\label{elBrm}%
\end{equation}
}\newline\newline becomes quite different.%

\begin{center}
\includegraphics[
trim=0.191207in 0.577709in 0.291859in 0.752846in,
height=2.0081in,
width=3.0113in
]%
{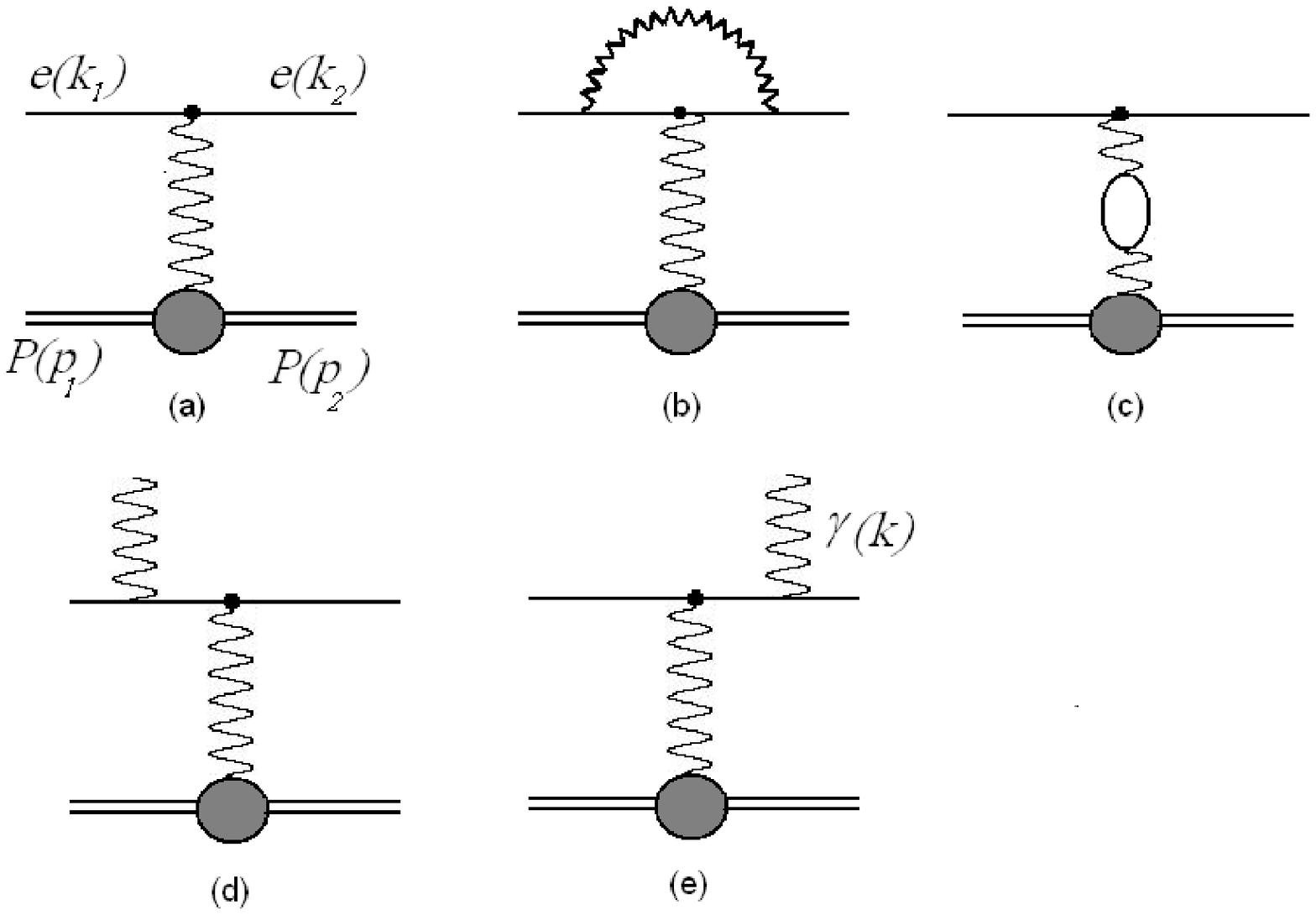}%
\\
\textbf{FIG. 1}{\small . Feynman diagrams contributing to the Born and
Radiative corrections cross sections.}%
\end{center}

{\normalsize The first comprehensive contribution in this direction has been
carried out by Akhundov et al.~\cite{Fortschr}. In that original review-paper,
the model-independent QED corrections in leptonic, hadronic, mixed and
Jaquet-Blondel variables were treated on a common base. Analytical and
semi-analytical formulae were derived for deep inelastic scattering~equation
(\ref{deepborn}). The QED radiative corrections for the process~(\ref{elborn})
have been recently calculated in the hadronic variables by considering
unpolarized and polarized parts of the cross sections \cite{Afanasev}.}

{\normalsize The main aim of this report is to present explicitly very compact
analytical formulae for QED radiative corrections to elastic $e(\mu)p $
scattering (see Fig.1) in the hadronic variables resulting from the general
approach of Akhundov et al.~\cite{Fortschr}. For this purpose we calculate the
QED corrections for the energies of JLab~\cite{JLab} using a new
parametrization of the form factors~\cite{Bosted}, and compare our numerical
results with corresponding ones of ref.~\cite{Afanasev}. }

{\normalsize \bigskip}

{\normalsize For the lepton-proton reactions (\ref{deepborn}) and
(\ref{elborn}), with the competing processes (3), the hadronic variables are
defined as follows \cite{Fortschr}:}%

\end{subequations}
\begin{equation}
Q_{h}^{2}=(p_{2}-p_{1})^{2},\text{ \ \ }y_{h}=\frac{p_{1}(p_{2}-p_{1})}%
{p_{1}k_{1}},\text{ \ \ \ }x_{h}=\frac{Q_{h}^{2}}{y_{h}S},\label{wav7}%
\end{equation}

{\normalsize where }%

\begin{equation}
s=-(k_{1}+p_{1})^{2}=S+m^{2}+M^{2},\label{wav8}%
\end{equation}

{\normalsize $m$ and $M$ are the masses of the incident lepton, and the proton
respectively. }

{\normalsize For the elastic scattering $x_{h}=1$, and
\begin{equation}
{Q_{h}}^{2}=Sy_{h}.\label{QSy}%
\end{equation}
}

{\normalsize The Born cross section of the process~(\ref{elborn}) in the
hadronic variables takes the form:}%

\begin{equation}
\frac{d\sigma^{\mathrm{B}}}{dQ_{h}^{2}}=\frac{2\pi\alpha^{2}}{{\lambda_{S}}%
}\sum_{i=1}^{3}\mathcal{A}_{i}(Q_{h}^{2})\frac{1}{Q_{h}^{4}}\;\mathcal{S}%
_{i}^{\mathrm{B}}(y_{h}),\label{cross}%
\end{equation}
{\normalsize \ where $\lambda_{S}=S^{2}-4m^{2}M^{2}$. \ The leptonic functions
$\mathcal{S}_{i}^{\mathrm{B}}$ appearing in equation (\ref{cross}) are:
\begin{align}
\mathcal{S}_{1}^{\mathrm{B}}(y_{h})  &  =Q_{h}^{2}-2m^{2},\\
\mathcal{S}_{2}^{\mathrm{B}}(y_{h})  &  =2[(1-y_{h})S^{2}-M^{2}Q_{h}^{2}],\\
\mathcal{S}_{3}^{\mathrm{B}}(y_{h})  &  =2Q_{h}^{2}(2-y_{h})S.\label{eqS3B}%
\end{align}
\qquad}

{\normalsize The hadronic functions $\mathcal{A}_{i}(Q_{h}^{2})$ in equation
(\ref{cross}) describe the electroweak interactions of leptons via the
exchange of a photon or $Z$ boson with unpolarized protons and are the
generalized elastic form factors of the nucleon. These form factors were
defined by formulae (20)-(22) of~Akhundov et al.\cite{ERT}.}

{\normalsize Using the results of ref.~\cite{Fortschr} (formulae (7.35),
(1.10) and (1.11)) we obtain, for the QED corrected cross section of the
elastic scattering~process (\ref{elborn}), the following expression: }

{\normalsize
\begin{equation}
\frac{d\sigma}{dQ_{h}^{2}}=\frac{d\sigma_{\mathrm{B}}}{dQ_{h}^{2}}\exp\left[
\frac{\alpha}{\pi}\delta^{\mathrm{inf}}(Q_{h}^{2})\right]  +\frac{2\alpha^{3}%
}{S^{2}}\sum_{i=1}^{3}A_{i}(Q_{h}^{2})\frac{1}{Q_{h}^{4}}\mathcal{S}_{i}%
(y_{h}).\label{d2sh}%
\end{equation}
\qquad}

{\normalsize The exponentiated dilogarithmic term $\delta^{\mathrm{inf}}%
(Q_{h}^{2})$ takes into account the vertex corrections (Fig.1b) and the
multiple soft photon emission:
\[
\delta^{\mathrm{inf}}(Q_{h}^{2})=\left(  \text{L}_{\text{h}}-1\right)
\ln(1-y_{h}),
\]
\qquad\qquad}

{\normalsize where }L$_{\text{h}}${\normalsize $=\ln(Q_{h}^{2}/m^{2})$. The
second term in equation~(\ref{d2sh}) is the hard bremsstrahlung correction of
order {$\mathcal{O}({\alpha})$} resulting from the threefold analytical
integration over the phase space of the process~(\ref{elBrm}) with emission of
the hard photon from the lepton (Fig.1d,1e): }%

\begin{align}
\mathcal{S}_{1}(y_{h})  &  =Q_{h}^{2}[\frac{1}{4}\ln^{2}y_{h}-\frac{1}{2}%
\ln^{2}(1-y_{h})-\ln y_{h}\ln(1-y_{h})\allowbreak\nonumber\\
&  -\frac{3}{2}\text{Li}_{\text{2}}(y_{h})+\frac{1}{2}\text{Li}_{\text{2}%
}(1)-\frac{1}{2}\ln y_{h}\text{L}_{\text{h}}\nonumber\\
&  +\frac{1}{4}\left(  1+\frac{2}{y_{h}}\right)  \text{L}_{\text{h1}%
}\allowbreak+\left(  1-\frac{1}{y_{h}}\right)  \ln y_{h}\nonumber\\
&  -\left(  1+\frac{1}{4y_{h}}\right)  ],\label{s1}%
\end{align}
\newline%

\begin{align}
\mathcal{S}_{2}(y_{h})  &  =S^{2}[-\frac{1}{2}y_{h}\ln^{2}y_{h}-(1-y_{h}%
)\ln^{2}(1-y_{h})\nonumber\\
&  -2(1-y_{h})\ln y_{h}\ln(1-y_{h})-y_{h}\text{Li}_{\text{2}}(1)\nonumber\\
&  -(2-3y_{h})\text{Li}_{\text{2}}(y_{h})+y_{h}\ln y_{h}\text{L}_{\text{h}%
}\nonumber\\
&  +\frac{y_{h}}{2}(1-y_{h})\text{L}_{\text{h1}}-\frac{y_{h}}{2}(2-y_{h})\ln
y_{h}],\label{s2}%
\end{align}

\begin{align}
\mathcal{S}_{3}(y_{h})  &  =SQ_{h}^{2}\{-(2-y_{h})[-\frac{1}{2}\ln^{2}%
y_{h}+\ln^{2}(1-y_{h})\nonumber\\
&  +2\ln y_{h}\ln(1-y_{h})+3\text{Li}_{\text{2}}(y_{h})-\text{Li}_{\text{2}%
}(1)\nonumber\\
&  +\ln y_{h}+\ln y_{h}\text{L}_{\text{h}}]+\frac{3}{2}y_{h}\text{L}%
_{\text{h1}}+y_{h}\nonumber\\
&  -\frac{7}{2}+2~(1-2y_{h})\ln y_{h}\},\label{s3}%
\end{align}

{\normalsize where
\begin{equation}
\text{L}_{\text{h1}}=\ln{\left(  \frac{Q_{h}^{2}}{m^{2}}\frac{y_{h}}{1-y_{h}%
}\right)  }.\label{wave9}%
\end{equation}
}

The formulae ({\normalsize \ref{s1})-(\ref{s3}) are ultra-relativistic, i.e.
}${\normalsize S>>M}^{2}.$ {\normalsize The running of the QED coupling
$\alpha$ may be taken into account in the standard way, as presented in refs.
\cite{Fortschr,Burhard}. }

At low energies the dominant contribution to ({\normalsize \ref{d2sh})} comes
from diagrams with $\gamma$ exchange, and all numerical calculations can be
performed using only the leptonic functions $S_{1,2}^{B}(y_{h})$,
$S_{1,2}(y_{h})$ and the form factors $A_{1,2}(Q_{h}^{2})$:%

\begin{equation}
A_{1}(Q_{h}^{2})=Q_{h}^{2}G_{M}^{2}(Q_{h}^{2}),\label{FormF1}%
\end{equation}%
\begin{equation}
A_{2}(Q_{h}^{2})=\frac{G_{E}^{2}(Q_{h}^{2})+\tau G_{M}^{2}(Q_{h}^{2})}{1+\tau
},\label{FormF2}%
\end{equation}
\newline\newline where $G_{E}^{2}(Q_{h}^{2})$ and $G_{M}^{2}(Q_{h}^{2})$ are
the standard electromagnetic form factors and $\tau=Q_{h}^{2}/4M^{2}$.%

\begin{center}
\includegraphics[
height=4.2203in,
width=3.1315in
]%
{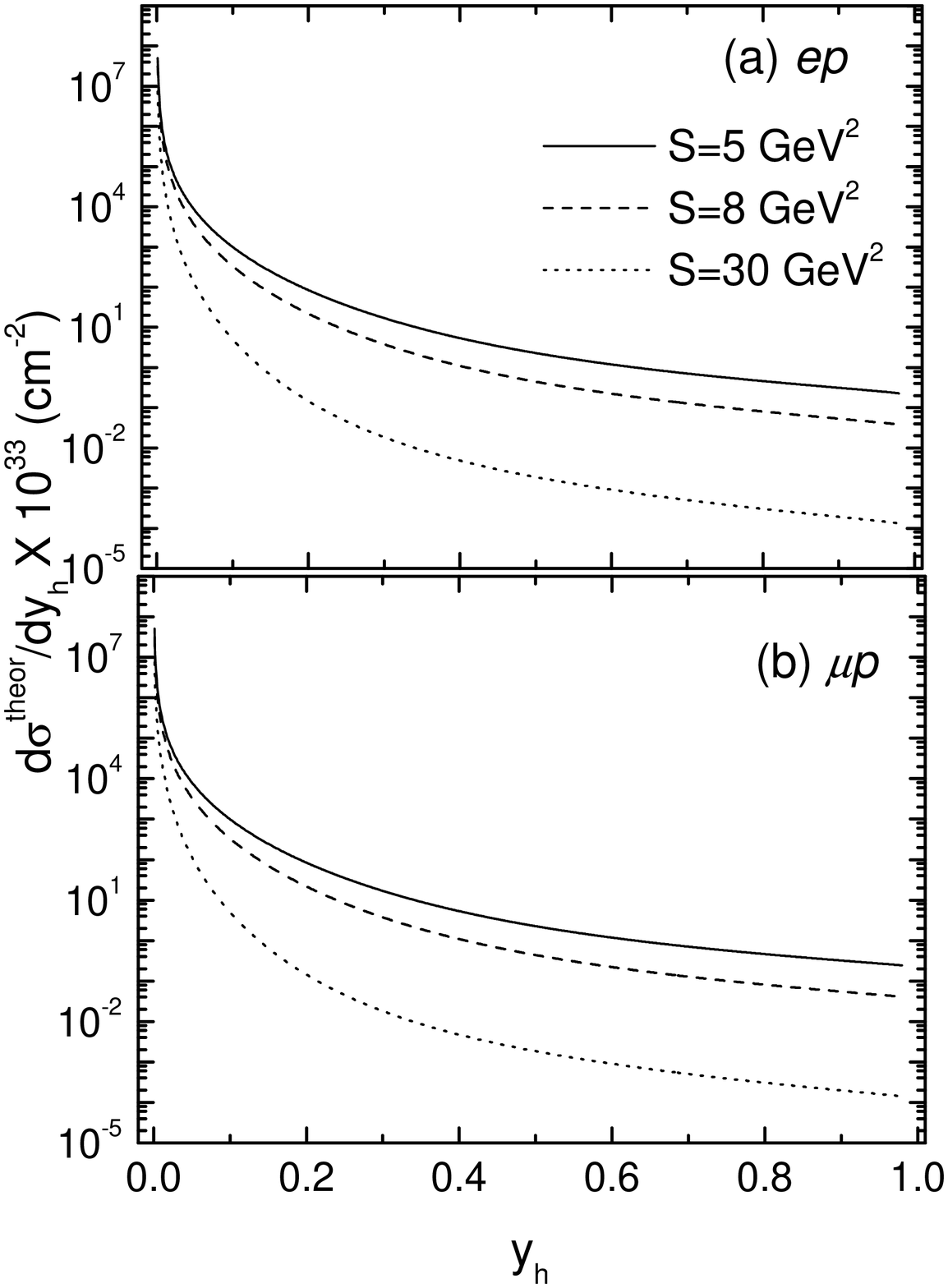}%
\\
\textbf{FIG.2.}{\small \ Cross-sections }$d\sigma/dy_{h}${\small , BORN+QED
corrections, at }$S=5,8,30$ {\small [}$GeV^{2}]${\small \ for }$ep$%
{\small scattering (a) and }$\mu p${\small scattering (b).}%
\end{center}

{\normalsize Figures 2 show the cross sections of $e(\mu)p$ elastic scattering
including the QED corrections for }${\normalsize S=5,8,30}${\normalsize \ }%
${\normalsize GeV}^{2}$. In our numerical calculations we have used the
parameterization of the form factors $G_{E}\left(  Q_{h}^{2}\right)  $ and
$G_{M}\left(  Q_{h}^{2}\right)  $ developed by Bosted
{\normalsize \cite{Bosted}.}

To compare our results with others, we use {\normalsize the radiative
correction factor $\delta(Q_{h}^{2})$ defined by:
\begin{equation}
\delta({Q_{h}^{2}})=\frac{d{\sigma}^{\mathrm{theor}}/{dQ_{h}^{2}}}{d{\sigma
}^{\mathrm{B}}/{dQ_{h}^{2}}}-1,\label{delta}%
\end{equation}
\qquad\qquad}

{\normalsize where $d{\sigma}^{\mathrm{theor}}/{dQ_{h}^{2}}$ is the
theoretical approximation to the measured cross section%
\begin{tabular}
[c]{c}%
$d{\sigma}^{\mathrm{meas}}$ $/$ ${dQ_{h}^{2}}$%
\end{tabular}
. }

{\normalsize The cross section~equation (\ref{d2sh}) and the contribution from
vacuum polarization (Fig. 1c) have been included in $d{\sigma}^{\mathrm{theor}%
}/{dQ_{h}^{2}}$. The running of the coupling }$\alpha$ has been found to
{\normalsize give a contribution of about }$3\%$ {\normalsize to the radiative
correction factor }$\delta.$

Figures 3 and 4 show $\delta\left(  Q_{h}\right)  $ for $S$=5, 8, 30 GeV$^{2}$
from our calculations (solid line) and from the calculations using formulae of
Afanasev et al.\footnote[1]{The formulae of Born cross section in the first
paper {\normalsize \cite{Afanasev}}\ contain an extra factor of 2.} (dashed
line) that contain the proton mass. As can be seen from the Figures 3 and 4,
we have a very good agreement in the order and the shape of radiative corrections.

For precise comparison between the two independent calculations we have
neglected the mass of proton in the formulae of Afanasev et al. (dotted lines
on Figures 3 and 4) and found a good agreement.

\begin{acknowledgments}
Two of us A.A. and H.A. would like to thank the research center at KSU for
support under project Phys/1420/27, and one of us H.H. would like to thank the
NCMP at KACST.
\end{acknowledgments}

\begin{tabular}
[c]{p{2.75in}cp{2.75in}}%
\multicolumn{1}{c}{%
{\parbox[b]{3.1315in}{\begin{center}
\includegraphics[
height=4.2203in,
width=3.1315in
]%
{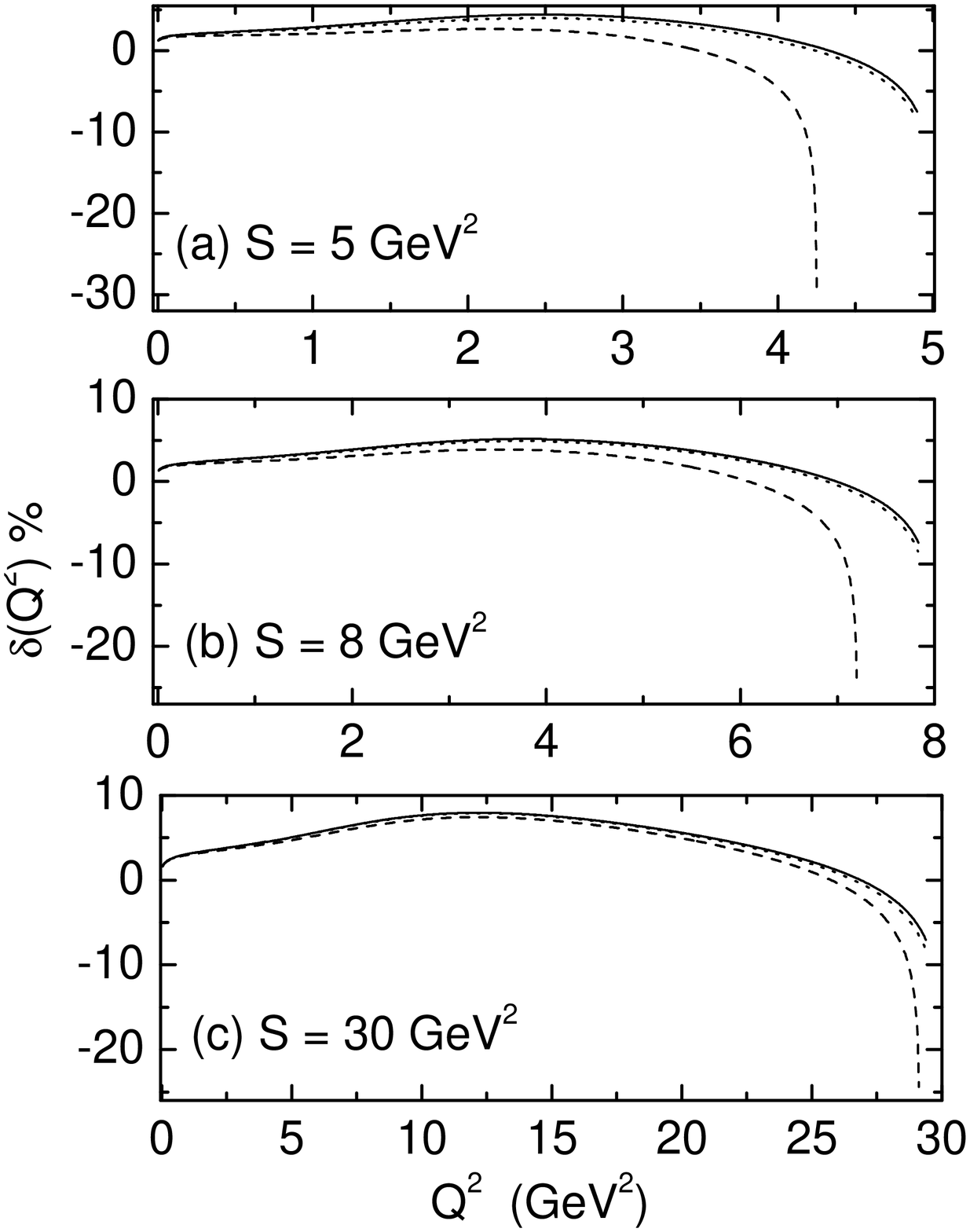}%
\\
{}%
\end{center}}}%
} &  & \multicolumn{1}{c}{%
{\parbox[b]{3.1315in}{\begin{center}
\includegraphics[
height=4.2203in,
width=3.1315in
]%
{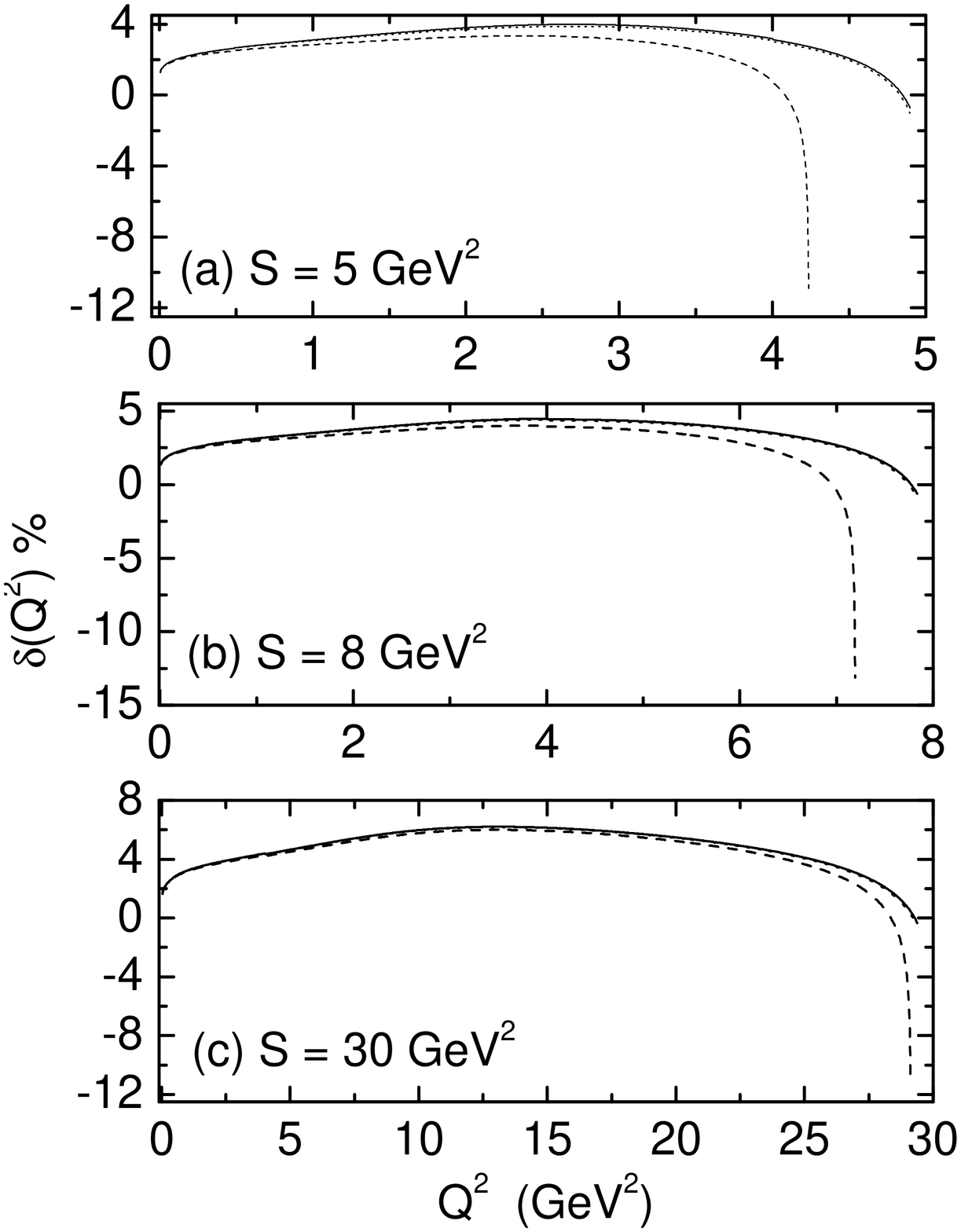}%
\\
{}%
\end{center}}}%
}\\
{\small FIG. 3. Radiative correction factor }$\delta\left(  Q_{h}^{2}\right)
${\small for }$ep${\small -scatering at (a) S=5 GeV}$^{2}${\small , (b) S=8
GeV}$^{2}${\small and (c) S=30 GeV}$^{2}${\small . Solid lines represent our
calculations and dashed and dotted lines are as calculated from the formulae
of ref. \cite{Afanasev}.} &  & {\small FIG. 4. Radiative correction factor
}$\delta\left(  Q_{h}^{2}\right)  $ {\small for }$\mu p${\small -scatering at
(a) S=5 GeV}$^{2}${\small , (b) S=8 GeV}$^{2}${\small and (c) S=30 GeV}$^{2}%
${\small . Solid lines are the present calculations and dashed and dotted
lines are as calculated from the formulae of ref. \cite{Afanasev}.}%
\end{tabular}

\end{document}